\documentclass[conference]{IEEEtran}
\IEEEoverridecommandlockouts
\usepackage[draft]{hyperref}
\usepackage{url}        
\usepackage{amsmath}
\usepackage{booktabs}     
\usepackage{amsfonts,amsmath,amssymb,amsthm, bbm, mathrsfs}      
\usepackage{nicefrac}       
\usepackage{microtype}      
\usepackage{lipsum}     
\usepackage{graphicx,subcaption}
\usepackage[numbers,sort&compress]{natbib} 
\usepackage{doi}
\usepackage{array}
\usepackage{enumitem}
\usepackage{bm}
\usepackage{color}
\usepackage{float}
\usepackage[dvipsnames]{xcolor}
\usepackage{multirow}
\usepackage[most]{tcolorbox} 
\usepackage{tikz}
\usepackage{makecell}
\usepackage[linesnumbered,ruled,vlined]{algorithm2e}
\usetikzlibrary{positioning}
\usepackage{algpseudocode}
\usepackage{soul}
\usepackage[most]{tcolorbox}
\usepackage{xcolor}

\def\BibTeX{{\rm B\kern-.05em{\sc i\kern-.025em b}\kern-.08em
    T\kern-.1667em\lower.7ex\hbox{E}\kern-.125emX}}

{
\theoremstyle{plain}
\newtheorem{keyquestion}{\textbf{Key Question}}

\newtheorem{definition}{\textbf{Definition}}
\newtheorem{theorem}{\textbf{Theorem}}

\newtheorem{lemma}{\textbf{Lemma}}

\newtheorem{assumption}{\textbf{Assumption}}
\newtheorem{remark}{\textbf{Remark}}

}

\newcommand{\epf}{{{e}}}

\newcommand{\toc}{{\varsigma_t}}
\newcommand{\gist}{{\textsc{gist1m}}}
\newcommand{\sift}{{\textsc{sift1m}}}
\newcommand{\msong}{{\textsc{msong}}}
\newcommand{\glove}{{\textsc{glove1.2m}}}

\def\bi{\begin{itemize}}
\def\bbi{\begin{itemize}}
\def\ei{\end{itemize}}
\def\eei{\end{itemize}}
\def\bc{\begin{center}}
\def\ec{\end{center}}

\begin{document}

\title{Trading Vector Data in Vector Databases}
\author{
    \IEEEauthorblockN{Jin Cheng\textsuperscript{$1,2$}, Xiangxiang Dai\textsuperscript{$2$},
    Ningning Ding\textsuperscript{$3$}, John C.S. Lui\textsuperscript{$2$}, Jianwei Huang\textsuperscript{$1*$}}
    \IEEEauthorblockA{
        \textsuperscript{$1$} The Chinese University of Hong Kong, Shenzhen \\
        \textsuperscript{$2$} The Chinese University of Hong Kong \\
        \textsuperscript{$3$} Hong Kong University of Science and Technology (Guangzhou) 
    }
}
\maketitle

\begin{abstract}
Vector data trading is essential for cross-domain learning with vector databases, yet it remains largely unexplored. We study this problem under online learning, where sellers face uncertain retrieval costs and buyers provide stochastic feedback to posted prices. Three main challenges arise: (1) heterogeneous and partial feedback in configuration learning, (2) variable and complex feedback in pricing learning, and (3) inherent coupling between configuration and pricing decisions.  

We propose a hierarchical bandit framework that jointly optimizes retrieval configurations and pricing. Stage I employs contextual clustering with confidence-based exploration to learn effective configurations with logarithmic regret. Stage II adopts interval-based price selection with local Taylor approximation to estimate buyer responses and achieve sublinear regret. We establish theoretical guarantees with polynomial time complexity and validate the framework on four real-world datasets, demonstrating consistent improvements in cumulative reward and regret reduction compared with existing methods.  

\end{abstract}

\section{Introduction}

\subsection{Background and Motivation}

Data has become a vital asset for data-driven technologies, particularly in machine learning and artificial intelligence (AI) \cite{cong2022data}. Access to diverse and large-scale datasets is essential for enhancing model robustness and generalization, enabling breakthroughs in applications such as large language models (LLMs) and recommendation systems \cite{gao2024causal}. To facilitate large-scale data sharing, the data trading market has experienced rapid growth. In 2022, the global market reached \$968 million and is projected to grow at an annual rate of 25\% from 2023 to 2030 \cite{datamarket2023}. This growth has been driven by the emergence of platforms such as AWS Data Exchange \cite{awsdata}, Snowflake \cite{snowflake}, and Xignite \cite{xignite}, which provide curated datasets to meet the growing demand for high-quality data in AI development.

\begin{figure}[hbpt]
    \centering
    \includegraphics[width=1\linewidth]{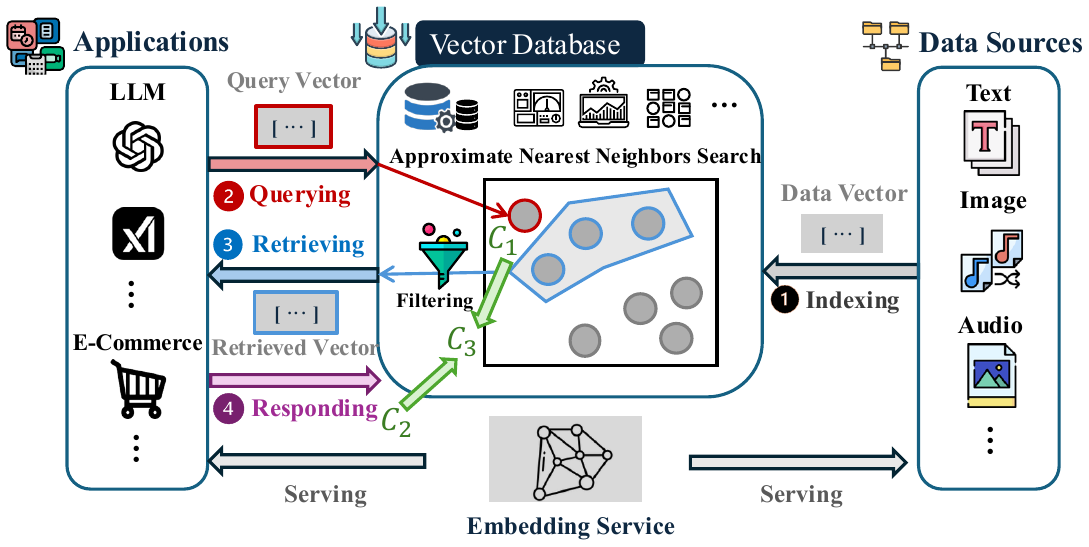}
    \caption{{Similarity-based retrieval from vector databases. }}
    \label{fig:intro:vdb}
    
\end{figure}

While traditional data trading platforms focus on structured data, vector data is essential for representing unstructured information and enabling efficient semantic retrieval in many AI tasks. In LLMs, vector databases are often integrated to support retrieval-augmented generation (RAG) \cite{pan2024survey}, where prompts and responses are enriched with semantically similar vectors. This mechanism mitigates hallucination, a key limitation of LLMs, by grounding responses in retrieved evidence \cite{huang2025survey}. {As illustrated in Fig.~}\ref{fig:intro:vdb}{, unstructured data such as text, images, or audio is embedded into vectors, then stored and indexed in vector databases. Given a query vector, an Approximate Nearest Neighbor (ANN) search retrieves the most similar vectors. An illustrative trading process is as follows: the buyer issues a query, the seller selects a retrieval configuration (e.g., indexing method, search depth) and sets a price, the system executes the query and returns approximate results, and the buyer simultaneously pays and provides a satisfaction signal reflecting relevance and cost-effectiveness, completing the transaction.}

{This workflow enables effective trading of embedding-based data across domains. In healthcare, hospitals may share medical image embeddings with researchers, where retrieval settings and pricing govern access to diagnostic data. In IoT, smart-city platforms may trade sensor embeddings to query patterns such as traffic or air quality, with sellers adjusting retrieval and pricing by system load and cost. In energy, organizations may exchange consumption or generation embeddings to support demand forecasting and grid optimization.}

However, research on trading vector data remains largely underexplored despite its growing importance in modern AI systems. Most existing data trading frameworks (e.g., \cite{hou2023scalable, chen2024gshop, wu2023variance, shi2024data, cai2023towards, chen2024efficient}) focus on structured data in traditional relational databases, where queries are executed over predefined tables with fixed schemas. These frameworks emphasize desirable properties for trading structured relational data, such as arbitrage-freeness~\cite{hou2023scalable,chen2024gshop}, fairness~\cite{wu2023variance,shi2024data}, and truthfulness~\cite{cai2023towards,chen2024efficient}.

{Vector data trading differs from structured-data markets in two main aspects: retrieval cost is uncertain due to high-dimensional geometry, query-dependent variability, and runtime stochasticity, and buyer responses are uncertain since utility is revealed only after approximate results and depends on semantic and task-specific factors. Such uncertainty in vector data trading necessitates an \emph{online learning} approach that adaptively optimizes retrieval configurations and pricing strategies based on \emph{interactive seller-buyer feedback}. However, online learning in this context introduces three key theoretical challenges.
}

{The first challenge ($C_1$ in Fig.~}\ref{fig:intro:vdb}{) is to adaptively optimize retrieval configurations under uncertain retrieval costs with \emph{heterogeneous and partial feedback}. Query diversity creates \emph{heterogeneity}: different queries favor different configurations, and the same configuration may yield varying outcomes due to system stochasticity. Feedback is also \emph{partial}, since only the chosen configuration is observed while alternatives remain untested. For example, during the retrieval process in Fig.~}\ref{fig:intro:vdb}{, the system applies one configuration and observes feedback only for that choice. This raises the first key question:}

\begin{keyquestion}
How to adaptively optimize retrieval configurations under uncertain retrieval costs with heterogeneous and partial feedback?
\end{keyquestion}

{The second challenge ($C_2$ in Fig.~}\ref{fig:intro:vdb}{) concerns optimizing pricing strategies under uncertain buyer responses with \emph{variable and complex feedback}. A buyer’s utility varies with query semantics, retrieval settings, and prices, producing highly \emph{variable} responses that are difficult to predict. Moreover, the perceived utility of retrieved results can affect future engagement, making responses \emph{complex}, often nonlinear and sensitive to small changes. This raises the second key question:}

\begin{keyquestion}
How to dynamically learn pricing strategies under uncertain buyer responses with variable and complex feedback?
\end{keyquestion}

{The third challenge ($C_3$ in Fig.~}\ref{fig:intro:vdb}{) arises from the \emph{coupling between retrieval and pricing}, as the two learning processes mutually affect each other. Retrieval configurations determine result quality, shaping buyers’ perceived utility and pricing responses, while pricing strategies influence query behavior, altering the distribution of query vectors and hence the learning of optimal configurations. This interdependence implies that optimizing them separately may be suboptimal. This raises the third key question:}

\begin{keyquestion}
How to jointly optimize retrieval configurations and pricing strategies under their inherent coupling?
\end{keyquestion}

\subsection{Key Contributions}

We summarize our key contributions as follows:

\begin{itemize}[leftmargin=*]
    \item \textbf{Framework for Trading Vector Data.}
    {To the best of our knowledge, this is the first work on vector data trading. We formulate a joint online learning problem over retrieval configurations and pricing strategies, and propose a hierarchical solution that addresses their inherent coupling (\emph{Key Question 3}) while generalizing to diverse retrieval backends.}
    
    \item \textbf{Two-Stage Hierarchical Bandit Algorithm.}  
    {We design a two-stage hierarchical bandit algorithm. Stage I (\emph{Key Question 1}) uses contextual clustering and confidence-based exploration to optimize retrieval configurations, while Stage II (\emph{Key Question 2}) learns dynamic pricing by identifying promising intervals and approximating buyer responses via local multivariate Taylor expansion.}

    \item \textbf{Theoretical Guarantees and Insights.}  
    {We establish regret bounds for both stages: near-optimal logarithmic regret for Stage I and sublinear regret for Stage II. The framework also operates with polynomial time complexity, ensuring theoretical soundness and computational efficiency.}

    \item \textbf{Experimental Validation.}  
    {Experiments on four real-world datasets show that our framework scales effectively and consistently outperforms baselines, achieving up to 29.9\% higher average reward and 87.7\% lower cumulative regret across diverse scenarios.}

\end{itemize}

The rest of the paper is organized as follows. Section \ref{sec:related} reviews the related works. Section~\ref{sec:system} introduces the system model. Section~\ref{sec:framework} presents the hierarchical bandit framework. Section~\ref{sec:stage1} and Section~\ref{sec:stage2} detail the learning algorithms for retrieval configuration and pricing, respectively. Section~\ref{sec:experiments} presents experimental results. Section~\ref{sec:conclusion} concludes the paper.

\section{Related Works} \label{sec:related}

We categorize related works into three areas: data trading methods (Section \ref{sec:relate:trade}), vector database optimization (Section \ref{sec:relate:database}), and multi-armed bandit frameworks (Section \ref{sec:relate:bandit}).

\subsection{Data Trading Methods} \label{sec:relate:trade}
The growing interest in data marketplaces has motivated extensive research on data trading methods. Existing studies mainly focus on structured relational data, developing trading mechanisms based on full information or game-theoretic models. These approaches typically rely on known buyer utilities or explicit preferences, and employ techniques such as Shapley-value-based allocation~\cite{wu2023variance,shi2024data}, budget-aware pricing~\cite{zhang2023survey,cai2022private}, and auctions~\cite{li2023multi,xinxin2023auction}. Researchers have also studied properties including privacy preservation~\cite{deng2025privacy,liu2024privacy,abla2024fair}, arbitrage-freeness~\cite{hou2023scalable,chen2024gshop}, truthfulness~\cite{cai2023towards,chen2024efficient}, and security~\cite{fang2024blockchain,xu2024secure,bauer2024designing}. Recent work extends these models to dynamic settings, such as online pricing with differential privacy~\cite{cai2021online}, incentive-compatible contracts~\cite{tian2019optimal}, and deterministic pricing for streaming queries~\cite{cheng2024continuous}. {To the best of our knowledge, existing data market mechanisms primarily focus on structured data, defined as data traded in its original form (e.g., relational tables, labeled records)~}\cite{zheng2017trading, huang2020fair, lu2024get}{. 
Some related works extend to unstructured sources by transforming them into derived features or aggregated statistics for collaborative learning~}\cite{li2020federated,hu2023data,zhou2022blockchain,kalra2023decentralized}{. 
Prior work assumes explicit utilities and targets data sharing or training, whereas vector data trading involves embeddings with uncertain utilities and configuration-dependent costs. }

\subsection{Vector Database Optimization}\label{sec:relate:database}
Vector databases are central to similarity-based retrieval in applications such as LLMs~\cite{pan2024survey} and recommendation systems~\cite{han2023comprehensive}. To support efficient large-scale search, they typically employ Approximate Nearest Neighbor (ANN) queries with indexing structures like the Inverted File Index (IVF)~\cite{jegou2010product} and Hierarchical Navigable Small World (HNSW) graphs~\cite{malkov2018efficient}. Recent work has enhanced graph-based indexing for dynamic and high-dimensional data, improving update efficiency~\cite{chen2023finger}, scalability~\cite{manohar2024parlayann}, hybrid queries~\cite{wang2023efficient}, and computational cost~\cite{gao2023high,zhao2023towards,niu2023residual}. Yet, these methods largely target data-level dynamics and are designed for static or batch query settings, offering limited support for adaptive configuration in online vector data trading, where retrieval must adjust continuously to uncertain market dynamics.

\subsection{Multi-armed Bandits Frameworks}\label{sec:relate:bandit}

Multi-armed bandit (MAB) algorithms are widely used for online decision-making, balancing exploration and exploitation under uncertainty. Classical methods such as UCB~\cite{auer2002finite} and Thompson Sampling~\cite{russo2018tutorial} are effective in stationary settings due to their simplicity. Advanced variants, including nonparametric~\cite{valko2013finite}, kernel-based~\cite{krause2008near}, Bayesian~\cite{zhu2023continuous}, and meta-learning approaches~\cite{khodak2023meta}, improve adaptability, while collaborative~\cite{dai2024conversational}, conversational~\cite{li2025towards}, federated~\cite{li2024fedconpe}, non-stationary~\cite{dai2024quantifying,dai2024axiomvision}, and multimodal-feedback methods~\cite{wang2018optimization, wang2021multimodal} address domain-specific challenges but not the coupled retrieval–pricing dynamics in vector data trading. Hierarchical frameworks~\cite{zuo2022hierarchical,baheri2025multilevel,liu2024analysis} support multi-stage decision-making but usually assume consistent feedback across stages. By contrast, vector data trading involves two tightly coupled stages with distinct feedback, requiring problem-aware and coordinated optimization.

This is the first work on vector data trading. To tackle its unique challenges, we propose a novel hierarchical bandit framework that jointly optimizes retrieval configurations and pricing through two coordinated stages. The next section introduces the system model that formally defines the problem.

\section{System Model}\label{sec:system}

In this section, we present the system model for vector data trading. We begin with the concept of \emph{vector queries} in Section~\ref{sec:sys:query}, which defines the trading products. Section~\ref{sec:sys:index} introduces the \emph{vector indexing} structures enabling efficient query retrieval. Section~\ref{sec:sys:trade} describes the \emph{online vector data trading} procedure, including buyer and seller modeling. Finally, Section~\ref{sec:sys:problem} formulates the online learning problem.

\subsection{Vector Queries}\label{sec:sys:query}

We focus on \emph{Approximate Nearest Neighbor (ANN)} queries, which are the trading products in our framework. These vector queries offer a practical trade-off between accuracy and efficiency, making them widely used in modern applications such as retrieval-augmented generation and recommendation systems \cite{genesis2025integrating}. {Following~}\cite{pan2024survey}{, we define an ANN query as:}

\begin{definition}[Approximate Nearest Neighbor (ANN) Query]
Given a query vector $\bm{v} \in \mathbb{R}^m$, an approximation factor $c > 1$, a retrieval size $k \in \mathbb{N}$, and a dataset $S \subset \mathbb{R}^m$, an \emph{Approximate Nearest Neighbor (ANN)} query $\bm{q} = (\bm{v}, c, k)$ returns a subset $S' \subseteq S$ of size $k$ such that:
\begin{equation}
    \forall \bm{x'} \in S', \quad d(\bm{x}', \bm{v}) \leq c \cdot \min_{\bm{x} \in S} d(\bm{x}, \bm{v}),
    \label{eq:def:ANN}
\end{equation}
where $d(\cdot, \cdot)$ denotes a distance metric (e.g., Euclidean).

\label{def:ANN}
\end{definition}

{An ANN query retrieves a set of vectors from a dataset $S \subset \mathbb{R}^m$ that are approximately similar to a query vector $\bm{v}$. Each returned vector must lie within $c \ge 1$ times the distance of the exact nearest neighbor, as given in Inequality (}\ref{eq:def:ANN}{). This multiplicative relaxation provides a practical balance between accuracy and efficiency in high-dimensional spaces, with $c$ typically controlled implicitly by system parameters.}

In addition to standard ANN queries, vector databases support a wide range of extended query types, including hybrid queries that combine vector similarity with attribute filtering, distance-bounded range queries, batch queries for parallel vector processing, and multi-vector aggregation queries~\cite{pan2024survey}. These query extensions complement ANN queries by improving the expressiveness of vector retrieval for downstream applications. Without loss of generality, we restrict our scope as follows:

\begin{remark}
For clarity and tractability, we focus on ANN queries, while other types such as hybrid, range, batch, and aggregation queries have been studied and can be integrated into our framework.
\end{remark}

We next review existing data trading methods for other query types and outline how they can be adapted to our framework. For \emph{hybrid queries} that combine vector similarity with attribute filtering, query pricing frameworks such as QIRANA~\cite{deep2017qirana} introduce attribute-aware pricing models, which can be layered onto our retrieval configuration without modifying its core structure. For \emph{range queries} that impose distance-based constraints, existing techniques \cite{cai2022private} compute prices based on query selectivity or coverage. These constraints can be incorporated into our pipeline as post-configuration filters. For \emph{batch queries} involving multiple vectors, recent studies employ bundling or workload-based optimization~\cite{niyato2016smart,zhang2023survey}, where our pricing module can act as a per-query subroutine within the broader batch pricing routine. Finally, for \emph{aggregation queries} such as similarity joins or representative queries, our hierarchical bandit framework remains applicable after suitable preprocessing. For example, the task can be decomposed into independent ANN queries, whose results and prices are recombined using established bundling strategies~\cite{niyato2016smart}.

\subsection{Vector Indexing}\label{sec:sys:index}

Vector indexing serves as a crucial bridge between vector queries and online data trading. It enables efficient execution of ANN queries by organizing high-dimensional data into structured and searchable forms. In our framework, we adopt a hybrid indexing scheme combining HNSW (Hierarchical Navigable Small World) and IVF (Inverted File Index), which is widely used in both academic research and production systems such as FAISS \cite{faiss}, Milvus \cite{milvus}, and Weaviate~\cite{weaviate}.

\subsubsection{\textbf{HNSW}}

The Hierarchical Navigable Small World (HNSW) indexing technique~\cite{malkov2018efficient} builds a multi-layer proximity graph that supports efficient navigation from coarse to fine granularity and achieves near-logarithmic search complexity. At each layer, nodes connect via greedy nearest neighbor search. Higher layers provide long-range links for global traversal, while lower layers support precise local search. During retrieval, HNSW performs a best-first search from the top layer down.

A key configuration parameter during retrieval is the \emph{expansion factor} $e$ (denoted as $ef$ in~\cite{malkov2018efficient}), which controls the size of the candidate set explored during the search. Although the approximation factor $c$ cannot be directly adjusted, $e$ serves as its practical proxy and governs the trade-off between retrieval quality and computational cost.

One key objective is to adaptively select the configuration parameter $e$, which directly controls retrieval cost and serves as a practical proxy for the approximation factor $c$. However, adaptive configuration of $e$ remains underexplored. Most systems use fixed configurations without adapting to individual queries, limiting their suitability in uncertain environments.

\subsubsection{\textbf{IVF}}

IVF (Inverted File Index) \cite{chowdhury2010introduction} is a widely used indexing method that complements HNSW by introducing coarse-grained clustering, reducing retrieval costs by filtering irrelevant candidates. It partitions the dataset into disjoint clusters, typically using $k$-means or similar methods, and restricts the search to clusters most relevant to the query vector.

Although IVF does not expose tunable parameters during retrieval, its clustering structure influences search behavior and pricing granularity. In our framework, IVF is used as a fixed preprocessing step, and the resulting clusters provide the basis for cluster-level learning and pricing, as detailed in Section~\ref{sec:stage1}.

\subsubsection{\textbf{Extensibility to Other Indexing Methods}}

Beyond HNSW and IVF, various indexing methods have been developed for approximate nearest neighbor search, such as tree-based (e.g., KD-tree \cite{men2025parallel}), hashing-based (e.g., LSH \cite{bawa2005lsh}), and quantization-based (e.g., PQ, OPQ \cite{ge2013optimized}) techniques. {More generally, when an indexing method involves multiple tunable parameters, our framework treats them jointly as a configuration vector and optimizes within the same hierarchical bandit process. In Stage I, the clustered contextual bandit operates over the higher-dimensional configuration space $\mathcal{E}$, where the theoretical analysis almost remains unchanged: the regret bound scales with its covering number but stays sublinear under standard assumptions. As discussed in Section V-B1, discretizing parameters into buckets keeps the space tractable, ensuring both flexibility and efficiency.}

\subsection{Online Vector Data Trading}\label{sec:sys:trade}
We next describe the interactive trading process between buyers and sellers, as illustrated in Fig.~\ref{fig:sys:datatrading}. The buyer and seller models are detailed in Sections \ref{sec:sys:trade:buyer} and~\ref{sec:sys:trade:seller}, respectively.\footnote{

{The framework functions as an external controller that interacts with vector databases through standard APIs.  
It sets the retrieval configuration $e_t$ (e.g., {nlist}, {nprobe} in IVF or $M$, $ef$ in HNSW) when issuing queries, and collects query performance and buyer feedback for learning.  
This design enables lightweight integration with systems such as FAISS and HNSWlib without modifying their internals.}}

\subsubsection{\textbf{Data Buyer Modeling}}\label{sec:sys:trade:buyer}
The buyer modeling comprises (a) the \emph{buyer–seller interaction} and (b) the \emph{buyer response}.

\begin{figure}
    \centering
    \includegraphics[width=0.95\linewidth]{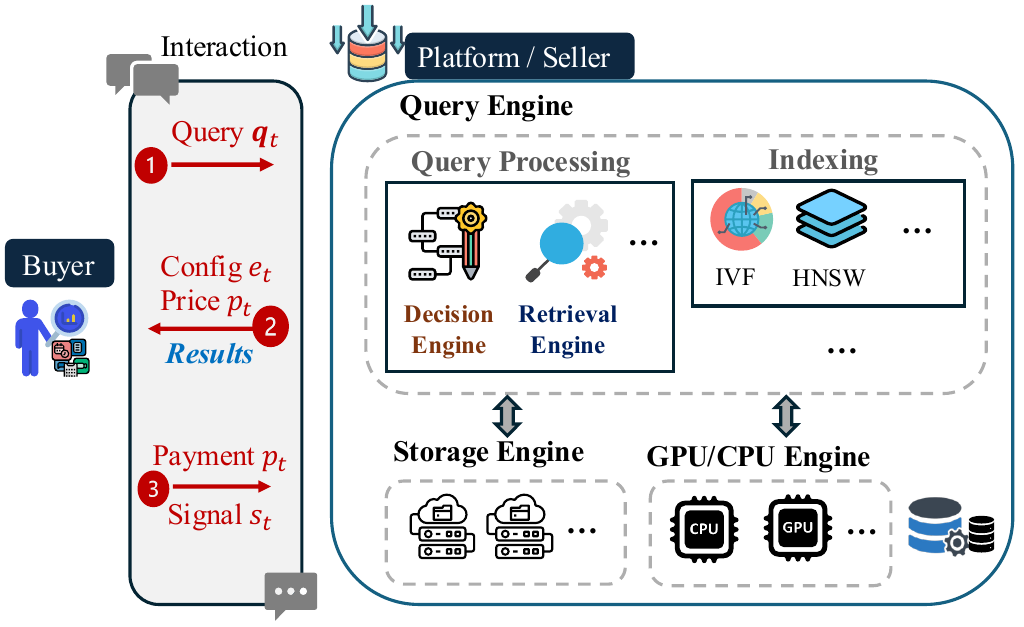}
    \caption{Vector-based Data trading system.}
    \label{fig:sys:datatrading}
\end{figure}

\textit{(a) \textbf{Buyer–Seller Interaction.}}
At each round $t \in \{1, 2, \ldots, T\}$, a data buyer submits an ANN query $\bm{q}_t = (\bm{v}_t, c_t, k_t)$ to the seller (i.e., the platform providing vector database services) (Step~\textcircled{1} in Fig.~\ref{fig:sys:datatrading}). The seller determines a retrieval configuration $e_t$ and a posted price $p_t$ (Step~\textcircled{2}), then executes the query using $e_t$ and returns the retrieved results along with the offer $(e_t, p_t)$. After observing both the results and the offer, the buyer pays the posted price $p_t$, with a response signal $s_t \in [0, 1]$ (Step~\textcircled{3}).

\textit{(b) \textbf{Buyer Response.}}
The response signal $s_t$, returned by the buyer in Step~\textcircled{3}, serves as the core feedback mechanism in the trading process. We model $s_t$ as a bounded random variable within $[0, 1]$, capturing the stochastic nature of buyer feedback. This signal plays a dual role: it reflects the buyer’s perceived utility from the current results and may influence future engagement behavior. First, $s_t$ quantifies the buyer’s utility from the retrieval results, shaped by both the retrieval configuration $e_t$, which determines the result quality, and the posted price $p_t$, which affects the perceived cost-effectiveness. Second, a higher $s_t$ indicates stronger buyer approval, potentially increasing the likelihood of future participation or repeated purchases.

However, modeling $s_t$ is challenging due to diverse query semantics and variations in structural features such as approximation level ($c_t$) and retrieval size ($k_t$), which together yield heterogeneous response patterns. A single global function may fail to capture this variability, leading to poor generalization.

To address this, we partition the query space into a finite set $\mathcal{C}$ of behaviorally similar clusters. A clustering function $h(\cdot)$ maps each query $\bm{q}_t$ to a cluster $\varsigma \in \mathcal{C}$ using both semantic and structural parameters (see Section~\ref{sec:stage1:design}). Each cluster runs an independent hierarchical bandit for configuration and pricing, enabling localized learning that adapts to diverse behaviors.

Based on the clustering structure, we define a response function $f^\varsigma: \mathcal{E} \times \mathcal{P} \to [0, 1]$ for each cluster $\varsigma \in \mathcal{C}$ to model the expected value of buyer response $s_t$ given the retrieval configuration and price. Here, $\mathcal{E}$ denotes the set of feasible configurations, and $\mathcal{P}$ represents the allowable price range $[\underline{p}, \bar{p}]$. The response function $f^{\varsigma}$ is defined as follows:

\begin{equation}
f^{\varsigma_t}(e_t, p_t)=\mathbb{E}[s_t \mid e_t , p_t ,\varsigma_t].
\label{eq:sys:buyer:satisfiction}
\end{equation}

\subsubsection{\textbf{Data Seller Modeling}}\label{sec:sys:trade:seller}

The seller denotes the platform offering vector database services and interacting with buyers. As shown in Fig.~\ref{fig:sys:datatrading}, it includes a decision engine for configurations and pricing, a retrieval engine for similarity search, and infrastructure for indexing (e.g., IVF, HNSW), storage, and compute. The design is compatible with real-world systems such as Milvus or FAISS on cloud infrastructure, and its learning-based decision module enables adaptive retrieval and pricing for monetizable vector data services.

In the following, we model the seller from two perspectives: (a) \emph{seller decision} and (b) \emph{seller reward}.

\textit{(a) \textbf{Seller Decision.}} At each round $t$, the seller selects a retrieval configuration $e_t \in \mathcal{E}$ and a posted price $p_t \in \mathcal{P}$, which jointly determine both the retrieval cost and the buyer’s response. The configuration $e_t$ governs the retrieval process and influences the quality of the returned results. Although buyers specify an approximation level $c_t$ in their queries, this parameter is often difficult to enforce directly in practice. Instead, the platform tunes the system-level parameter $e_t$ to indirectly control the effective approximation, thereby shaping the quality of the result set. For more details on the indexing structure and the role of $e_t$, refer to Section~\ref{sec:sys:index}.

\textit{(b) \textbf{Seller Reward.}}
{The seller’s reward reflects both buyer response and seller profit in each round. At round $t$, the seller pays an execution cost $c_t$ to process query $\bm{q}_t$ under configuration $e_t$ and receives payment $p_t$ from the buyer. Following prior work on data markets~}\cite{lu2024get,zheng2017trading,huang2020fair}{, the profit equals $(p_t - c_t)$. Incorporating the buyer’s response signal $s_t \in [0,1]$, the realized reward is:}
\begin{equation}
r_t = s_t \cdot (p_t - c_t).
\end{equation}

{This formulation links the reward to both profit and buyer response, capturing immediate gains and long-term engagement. We model the execution cost $c_t$ using a cluster-specific cost function $g^{\varsigma_t}(e_t)$, which quantifies the system resources required to process $\bm{q}_t$ in cluster $\varsigma_t$ under configuration $e_t$. The function can incorporate CPU usage, memory footprint, or query latency from profiling data, and may be instantiated as a weighted or nonlinear combination of multiple resource factors, or augmented with budget constraints as needed. Formally, for each cluster $\varsigma \in \mathcal{C}$, we define:}

\begin{equation}
g^{\varsigma_t}(e_t) = \mathbb{E}[c_t \mid e_t, \varsigma_t].
\label{eq:sys:g}
\end{equation}

Accordingly, we define the expected reward function as: \footnote{This formulation assumes conditional independence between the response signal $s_t$ and execution cost $c_t$. Here, $s_t$ captures buyer-perceived utility, while $c_t$ reflects backend resource usage. Under this assumption, the expectation decomposes as $\mathbb{E}[s_t \cdot (p_t - c_t)] = \mathbb{E}[s_t] \cdot (p_t - \mathbb{E}[c_t]).$
}

\begin{equation}
\begin{aligned}
 u_t^{\varsigma_t}(e_t, p_t)&  = f^{\varsigma_t}(e_t, p_t) \cdot \left(p_t - g^{\varsigma_t}(e_t)\right),
\end{aligned}
\label{eq:sys:sellerreward}
\end{equation}

\noindent where $f^{\varsigma}(e_t, p_t)$ and $g^{\varsigma_t}(e_t)$ are defined in Equations~(\ref{eq:sys:buyer:satisfiction}) and~(\ref{eq:sys:g}), respectively.

\subsection{Problem Formulation}\label{sec:sys:problem}

This work focuses on an {online learning optimization problem} that aims to maximize the seller’s cumulative expected reward over a time horizon $T$. To formalize this objective, we reformulate it as an \emph{online regret minimization problem} $\mathbb{P}_1$, which is equivalent to maximizing cumulative expected reward.

At each round $t$, the platform selects a retrieval configuration $e_t \in \mathcal{E}$ and a posted price $p_t \in \mathcal{P}$ for the query assigned to cluster $\varsigma_t$. This decision pair $(e_t, p_t)$ jointly determines the seller’s expected reward, as defined in Equation~(\ref{eq:sys:sellerreward}). To evaluate its effectiveness, we compare it against the optimal decision pair $(e^*, p^*)$ that would maximize the expected reward. The instantaneous regret $z_t^{\varsigma_t}(e_t, p_t)$  is defined as:
\begin{equation}
z_t^{\varsigma_t}(e_t, p_t) = u_t^{\varsigma_t}(e^*, p^*) - u_t^{\varsigma_t}(e_t, p_t).
\end{equation}

We define the cumulative regret for cluster $\varsigma$ as the sum of instantaneous regrets over all rounds assigned to it:

\begin{equation}
    R_\varsigma(T_\varsigma) = \sum_{t=1}^T \bm{1}(\varsigma_t = \varsigma) \cdot z_t^{\varsigma_t}(e_t, p_t),
\end{equation}

\noindent where $T_\varsigma$ denotes the number of rounds involving cluster $\varsigma$.\footnote{

{The regret can also be written as 
$R_T^{\mathrm{w}}=\sum_{\varsigma} w_{\varsigma} R_T^{\varsigma}$ 
with weights for traffic or variance. The cluster-based analysis still yields a sublinear bound under the same assumptions.}
}

Aggregating across all clusters, we formalize the online learning problem studied in this work as follows:

\noindent\textbf{Problem} $\mathbb{P}_1$: \emph{Online Regret Minimization}

\begin{equation}
    \min_{\bm{e},\bm{p}}~\sum_{\varsigma \in \mathcal{C}} R_\varsigma(T_\varsigma) \tag{$\mathbb{P}_1$}
\end{equation}
\begin{equation}
    \begin{aligned}
        \text{s.t.} \quad & e_t \in \mathcal{E}, ~p_t \in \mathcal{P}, & \forall t \in \{1,\cdots,T\}. \notag
    \end{aligned}
\end{equation}

Solving Problem $\mathbb{P}_1$ presents several \textbf{technical challenges}:

\begin{itemize}[leftmargin=*]

\item \emph{Heterogeneous and partial feedback in retrieval configurations (Key Question 1).}
The retrieval cost $c_t$ is uncertain due to system stochasticity, and the expected cost $g^\varsigma(e_t)$ must be learned online. Feedback is \emph{partial}, as only the chosen configuration $e_t$ is observed, and \emph{heterogeneous}, since different queries may favor different configurations.

\item \emph{Variable and complex feedback in pricing strategies (Key Question 2).}
The response signal $s_t$ varies with query semantics and retrieval outcomes, and its expectation $f^\varsigma(e_t,p_t)$ must be learned online. Such feedback is \emph{variable} and often \emph{complex} and nonlinear, creating challenges for optimization.

\item \emph{Inherent coupling between retrieval and pricing (Key Question 3).}
The reward $r_t$ depends jointly on $e_t$ and $p_t$. Higher $e_t$ improves quality but raises costs, requiring higher $p_t$ to stay profitable, while higher $p_t$ may reduce buyer satisfaction. This trade-off necessitates coordinated learning of $e_t$ and $p_t$.

\end{itemize}

In this section, we introduced the system model for vector data trading. The optimization problem $\mathbb{P}_1$ poses three technical challenges, motivating the need for a specialized learning framework. In the next section, we introduce a two-stage hierarchical bandit approach to address Problem~$\mathbb{P}_1$.

\section{Hierarchical Bandit Framework} \label{sec:framework}

We address Problem~$\mathbb{P}_1$ by developing a two-stage hierarchical bandit framework with theoretical guarantees. Section~\ref{sec:framework:overview} presents the framework overview, Section~\ref{sec:framework:design} describes the algorithmic design, and Section~\ref{sec:framework:regret} provides the regret decomposition that supports the analyses in Sections~\ref{sec:stage1} and~\ref{sec:stage2}.

\subsection{Framework Overview}\label{sec:framework:overview}

\begin{figure}[t]
    \centering
    \includegraphics[width=0.65\linewidth]{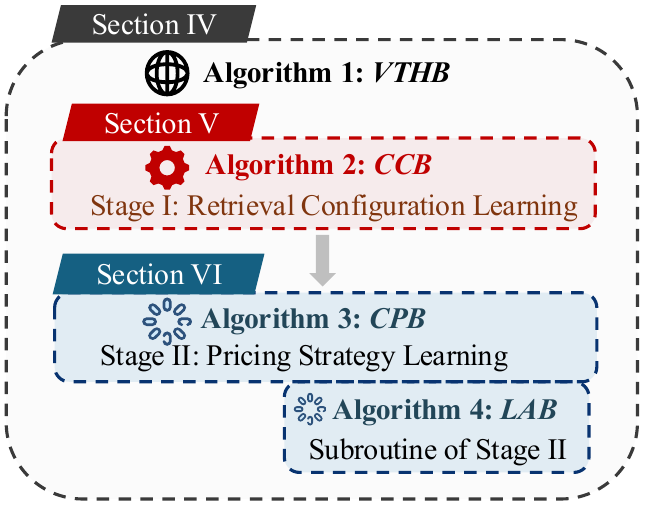}
    \caption{Framework Overview.}
    \label{fig:framework:overview}
    
\end{figure}

Fig. \ref{fig:framework:overview} presents an overview of the proposed framework, comprising four interrelated algorithms:

\begin{itemize}[leftmargin=*]
\item \textbf{Algorithm 1: \textit{VTHB}} (Section~\ref{sec:framework}) serves as the top-level controller that coordinates the entire online trading process.
\item \textbf{Algorithm 2: \textit{CCB}} (Section~\ref{sec:stage1}) handles retrieval configuration learning. It assigns queries to clusters and adaptively selects configurations using confidence-based exploration.

\item \textbf{Algorithm 3: \textit{CPB}} (Section~\ref{sec:stage2}) manages pricing strategy learning. It selects promising price intervals and determines the final price within the selected interval via Algorithm~4.

\item \textbf{Algorithm 4: \textit{LAB}} (Section~\ref{sec:stage2}) is a subroutine of \textit{CPB} that performs local optimization via Taylor approximation.

\end{itemize}

\begin{algorithm}[htbp]
\caption{Hierarchical Bandit for Trading Vector Data (\textit{VTHB})
}
\KwIn{$\mathcal{E}$, $\mathcal{P}$}
\textbf{Initialization}: historical data $\mathcal{H}=\emptyset$\;
\For{$t\in\{1,\cdots,T\}$}{
    Receive query $\bm{q_t}=(\bm{v_t},c_t,k_t)$\;

    $(\varsigma_t,e_t) \leftarrow \text{\textit{CCB}}(\bm{q_t}, \mathcal{E}, \mathcal{H})$\;

    $(j_t,p_t )\leftarrow \text{\textit{CPB}}(\varsigma_t, e_t, \mathcal{P}, \mathcal{H})$\;
    Execute $\bm{q}_t$ with configuration $e_t$, incurring cost $c_t$\;
    Return retrieved results together with $(e_t, p_t)$\;
    Observe buyer response signal $s_t$\;
    Compute reward $r_t = s_t\left(p_t  -c_t\right)$\;
    Compute utility $y_t=r_t/p_t$ \;
    Add $(\varsigma_t,e_t,j_t,p_t,s_t,r_t,y_t)$ to $\mathcal{H}$\;
}

\end{algorithm}

\subsection{Algorithm Design: VTHB}\label{sec:framework:design}

Algorithm~1 (\textit{VTHB})\footnote{
{We adopt a hierarchical bandit design that separates retrieval configuration and pricing, as these stages have distinct feedback and objectives. This decomposition ensures tractability and allows stage-specific learners. Joint optimization is possible but suffers from large action spaces and sparse signals, so we view it as a complementary extension when richer feedback is available.}} coordinates the online trading process over $T$ rounds. The platform initializes an empty history set $\mathcal{H}$ to record past interactions (Line 1). At each round $t$, a query $\bm{q}_t$ arrives with its retrieval requirements. (Lines 2–3).

In Stage I, the platform calls the \textit{CCB} algorithm (Line~4) to assign the query to a cluster $\varsigma_t$ and select a retrieval configuration $e_t$ based on contextual clustering and confidence-based exploration. In Stage II, the \textit{CPB} algorithm is invoked (Line~5) to identify a promising price interval $j_t$ and determine a posted price $p_t$ using the subroutine \textit{LAB}.

The platform then executes the query using the selected configuration, incurring a retrieval cost $c_t$ (Line~6), and returns the result along with the offer $(e_t, p_t)$ (Line~7). The buyer provides a response signal $s_t \in [0,1]$ indicating perceived utility and future engagement (Line~8). Based on this interaction, the platform computes the realized reward $r_t$ (Line~9) and a normalized utility $y_t$ (Line~10) to guide future pricing decisions. Finally, all variables from this round are recorded in $\mathcal{H}$ to support iterative learning (Line~11).

Before detailing the Stage I and II algorithms\footnote{
{Retrieval shapes buyer acceptance, and pricing depends on retrieval outcomes. Optimizing them sequentially preserves tractability while capturing their interdependence through cumulative reward.}
}, we present the regret decomposition of Problem~$\mathbb{P}_1$, which decouples the analyses in Sections~\ref{sec:stage1} and~\ref{sec:stage2}.

\subsection{Regret Decomposition}\label{sec:framework:regret}

The cluster regret $R_\varsigma(T_\varsigma)$ in Problem~$\mathbb{P}_1$ can be decomposed into configuration and pricing regret as follows:

\begin{equation}
R_\varsigma(T_\varsigma) = R_\varsigma^c(T_\varsigma) + R_\varsigma^p(T_\varsigma),
\label{eq:framework:decompose}
\end{equation}
where $R_\varsigma^c(T_\varsigma)$ denotes the regret from selecting suboptimal retrieval configuration, and $R_\varsigma^p(T_\varsigma)$ denotes the regret from posting suboptimal prices under the optimal configuration. These two components are handled by different learners, as detailed in Sections~\ref{sec:stage1} and~\ref{sec:stage2}.

This section introduced the hierarchical bandit framework and the regret decomposition of Problem~$\mathbb{P}_1$, which defines the optimization objectives for Sections \ref{sec:stage1} and \ref{sec:stage2}. The time complexity analysis will be presented in Theorem~\ref{thm:time} of Section~\ref{sec:stage2} after all algorithmic components are introduced.

\section{Stage I: Retrieval Configuration Learning} \label{sec:stage1}

In this section, we present the clustered configuration bandit algorithm (\textit{CCB}) for adaptively selecting retrieval configurations under heterogeneous and partial feedback in Stage I, aiming to minimize the configuration regret $R_\varsigma^c$ in Equation~(\ref{eq:framework:decompose}). Section~\ref{sec:stage1:setup} introduces the algorithm setup, Section~\ref{sec:stage1:design} presents the algorithm design, and Section~\ref{sec:stage1:analysis} provides the theoretical analysis, showing that \textit{CCB} achieves logarithmic regret with polynomial time complexity.

\subsection{Algorithm Setup}\label{sec:stage1:setup}

In Stage~I, the platform must adaptively select a retrieval configuration $e_t$ for each query $\bm{q}_t$. To address the challenges of heterogeneous and partial feedback, our approach integrates \emph{contextual clustering} with \emph{confidence-based exploration}.

To tackle the heterogeneity challenge, we partition incoming queries into contextually homogeneous \emph{clusters}. This design is motivated by the nature of vector databases, which operate in high-dimensional semantic spaces and serve diverse, large-scale query workloads. Retrieval behavior varies significantly across query types—some require high precision with tight approximation, while others favor efficiency over accuracy. Clustering narrows the learning scope within each cluster, enabling localized adaptation and improved statistical efficiency.

To handle the partial feedback challenge, we adopt a \emph{confidence}-based exploration strategy within each cluster. The platform maintains optimistic reward estimates for each configuration using the Upper Confidence Bound (UCB) principle \cite{auer2002finite}. This approach explores uncertain configurations while favoring those with strong performance.

By combining semantic clustering with principled exploration, our design addresses \emph{Key Question 1}. We now proceed to detail the design of Algorithm 2 (\textit{CCB}).

\subsection{Algorithm Design: CCB}\label{sec:stage1:design}

As shown in Algorithm~2, \textit{CCB} comprises two key components: a context clustering mechanism (Line 1), which enables localized learning across diverse query types (see Section~\ref{sec:stage1:design:cluster}), and a confidence-based exploration strategy (Lines 2–9), which adaptively selects retrieval configurations within each cluster (see Section \ref{sec:stage1:design:learning}).

\begin{algorithm}[t]
\caption{Clustered Configuration Bandit (\textit{CCB})}
\KwIn{$\bm{q}_t$, $\mathcal{E}$, $\mathcal{H}$}
Compute cluster $\varsigma_t = h(\bm{q_t})$\;
\For{each $e \in \mathcal{E}$}{
    Retrieve $\mathcal{R}_e=\{r_{t'}|{ e_{t'} = e, \varsigma_{t'}=\varsigma_t}\}$ from $\mathcal{H}$\;
    \If{$\left|\mathcal{R}_e\right|=0$}{$\gamma_e=\infty$}
    \Else{    Compute $\phi_e =\sqrt{\frac{2\log t}{\left|\mathcal{R}_e\right|}}$\;
    Compute $\gamma_e= \frac{1}{\left|\mathcal{R}_e\right|} \sum_{r\in\mathcal{R}_e} r_t +\phi_e$\;}
}
 Select $e_t=\arg\max_{e \in \mathcal{E}} \gamma_e$\;
\KwOut{($\varsigma_t,e_t$)}
\end{algorithm}

\subsubsection{\textbf{Context Clustering for Localized Learning}}\label{sec:stage1:design:cluster}

We cluster queries based on semantic and structural properties, enabling the platform to assign an independent bandit learner to each cluster for localized adaptation and efficient exploration. Each query $\bm{q}_t = (\bm{v}_t, c_t, k_t)$ is assigned to a cluster $\varsigma_t$ as follows:

\begin{equation}
\varsigma_t = h(\bm{q}_t) = \left( \mathrm{IVF}(\bm{v}_t),b_c(c_t),b_k(k_t) \right),
\end{equation}

where $\mathrm{IVF}(\bm{v}_t)$ denotes the semantic centroid derived from the Inverted File Index (IVF) (see Section~\ref{sec:sys:index}). $b_c(\cdot)$ and $b_k(\cdot)$ are logarithmic bucketing operations defined as: $b_c(c) = B_c - \lfloor \log_2 c \rfloor$ and $b_k(k) = \lfloor \log_2 k \rfloor.$

\begin{figure}[hbpt]
    \centering
    \includegraphics[width=0.25\textwidth]{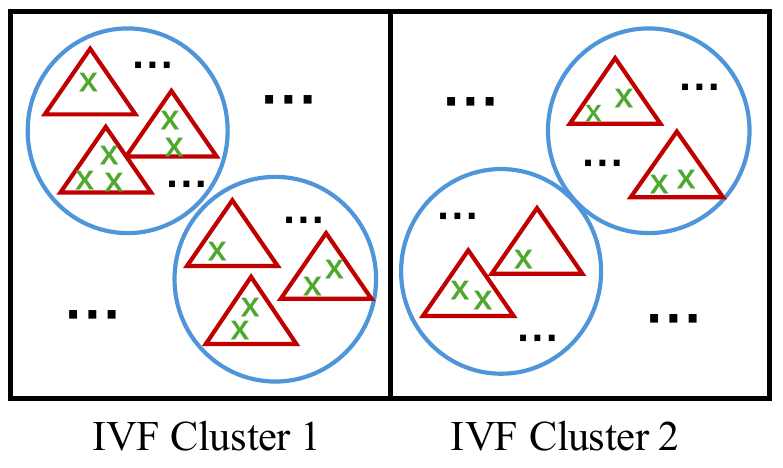}
    \caption{{Illustration of the clustering function $h(\cdot)$.}}
    \label{fig:clustering_h2}
    
\end{figure}

\noindent  The constant $B_c = \lceil \log_2 c_{\max} \rceil + 1$ is chosen to cover the full range of approximation factors, where $c_{\max}$ denotes the maximum approximation level. This design clusters queries with similar semantic and structural characteristics. {As illustrated in Fig.~}\ref{fig:clustering_h2}{, the number of clusters is determined by {nlist}, $B_c$, and $B_k$, which together balance granularity and sample stability. Rather than relying on explicit similarity thresholds, IVF centroids and logarithmic bucketing implicitly create meaningful partitions. Experiments in Section VII demonstrate that this design achieves stable rewards and lower intra-cluster variance.}\footnote{
{In our framework, clusters are defined by IVF partitioning and structural parameters ($c$, $k$), which remain fixed during the process. This design ensures sample sufficiency and tractable regret analysis.  
At the same time, query distributions may evolve. A natural extension is to adopt adaptive clustering, such as updating centroids online or refining clusters under drift, to better capture emerging patterns. We leave this as future work.}
}

\subsubsection{\textbf{Confidence-Based Exploration Strategy}}\label{sec:stage1:design:learning}

Once a query is assigned to a cluster, the platform selects a retrieval configuration $e_t \in \mathcal{E}$ using the Upper Confidence Bound (UCB) principle. For each configuration $e\in\mathcal{E}$, the platform maintains a history of observed rewards $\mathcal{R}_e$ within the current cluster (Line 3). At round~$t$, an optimistic reward estimate $\gamma_e$ is computed for each configuration $e \in \mathcal{E}$ (Lines~4–8), where
$\gamma_e = \infty$ if $|\mathcal{R}_e|=0$, and $\gamma_e = \widehat{r}_e + \phi_e$ otherwise.

Here, $\widehat{r}_e$ denotes the average historical reward of configuration $e$, and $\phi_e$ denotes the confidence radius (defined in Line~7), which quantifies the uncertainty due to limited feedback.

At round~$t$, the platform selects the configuration with the highest optimistic estimate (Line~9), favoring options with high past rewards or limited observations. This balances exploitation and exploration, refining estimates while remaining robust to noise and partial feedback.

By applying context clustering and confidence-based exploration, the platform achieves logarithmic regret and polynomial complexity, as detailed in the next subsection.

\subsection{Theoretical Analysis}\label{sec:stage1:analysis}

We analyze the efficiency of Algorithm~2 in terms of both time complexity and regret bounds. First, Algorithm~2 selects configurations by computing optimistic estimates $\gamma_e$ for all $e \in \mathcal{E}$ within the assigned cluster. With incremental updates, this step has a per-round time complexity of $O(|\mathcal{E}|)$.

We next analyze the configuration regret $R_\varsigma^c$, as decomposed in Equation~(\ref{eq:framework:decompose}) in Section~\ref{sec:framework:regret}, for Problem~$\mathbb{P}_1$ under two settings. Section~\ref{sec:stage1:analysis:special} considers the \emph{gap-dominant} case, where rewards are well-separated, enabling rapid identification of the optimal configuration. Section~\ref{sec:stage1:analysis:general} addresses the \emph{general} setting. In both cases, the cumulative regret is shown to grow logarithmically with the number of rounds.

\subsubsection{\textbf{Gap-Dominant Setting}}\label{sec:stage1:analysis:special}

We begin by analyzing a scenario where the reward distributions of different configurations are clearly separable, referred to as the \emph{gap-dominant reward structure}. We formalize this setting as follows:

\begin{definition}[Gap-Dominant Reward Structure]
For each cluster $\varsigma$, let $e^*$ denote the optimal configuration. The configuration setting exhibits a \emph{gap-dominant reward structure} if:
\begin{enumerate}[label=(\arabic*)]
\item For each $e \in \mathcal{E}$ in cluster $\varsigma$, there exists a fixed reward interval $[l_e^\varsigma, u_e^\varsigma]$ such that for all feasible prices, the expected reward $u^\varsigma_t(e, p) \in [l_e^\varsigma, u_e^\varsigma]$;
\item For any $e \ne e^*$, the intervals $[l_e^\varsigma, u_e^\varsigma]$ and $[l_{e^*}^\varsigma, u_{e^*}^\varsigma]$ are disjoint; i.e., $u_e^\varsigma < l_{e^*}^\varsigma$.
\end{enumerate}
\end{definition}

Here, $[l_e^\varsigma, u_e^\varsigma]$ represents the reward interval for the expected reward of configuration $e$ in cluster $\varsigma$ across all feasible prices. The disjointness assumption ensures a fixed margin between the optimal and any suboptimal configuration, simplifying the learning process. Let $\delta_e^\varsigma = l_{e^*}^\varsigma - u_e^\varsigma > 0$ denote the minimal reward gap between $e^*$ and a suboptimal configuration $e$ within cluster $\varsigma$ and let $\bar{r}$ represent the upper bound of $r_t$. Under this structure, the following theorem holds:

\begin{theorem}[Regret Bound of \textit{CCB} under Gap-Dominant Reward Structure]
\label{thm:gap-regret}
Suppose the configuration setting satisfies the gap-dominant reward structure, the expected cumulative configuration regret $ R^{c}_{\varsigma}(T_{\varsigma}) $ of Algorithm 2 over $T_{\varsigma}$ rounds is bounded by:

\begin{equation}
R^{c}_{\varsigma}(T_{\varsigma})\le\bar{r}
\sum_{e \ne e^*} \left( 3 + \frac{8 \log T_{\varsigma}}{(\delta_e^\varsigma)^2} + 2 \log T_{\varsigma} \right).
\end{equation}
\end{theorem}

The proof of Theorem 1 is given in Appendix A in the technical report \cite{report}. This theorem highlights that the configuration learner achieves logarithmic regret with respect to the number of rounds $T_\varsigma$, where the dominant term \( \frac{8 \log T_\varsigma}{(\delta_e^\varsigma)^2} \) captures the exploration cost. A larger separation gap $\delta_e^\varsigma$ leads to faster convergence by making suboptimal configurations easier to eliminate.

\subsubsection{\textbf{General Setting}}\label{sec:stage1:analysis:general}

We now consider the general case in which reward intervals of different configurations may overlap. Let $\hat{r}_e^\varsigma(n)$ be the empirical mean reward for configuration $e$ after $n$ selections within cluster $\varsigma$. As Stage II achieves sublinear regret for pricing learning (as shown in Section\ref{sec:stage2:analysis}), the empirical mean reward $ \mathbb{E}[\hat{r}_e^\varsigma(n)]$ converges to the true expected reward $u_t^{\varsigma}(e,p^*)$ as $n$ increases, as follows:

\begin{equation}
\lim_{n \to \infty} \mathbb{E}[\hat{r}_e^\varsigma(n)] = u_t^{\varsigma}(e,p^*),
\label{eq:stage1:er}
\end{equation}

\noindent where $p^*$ denotes the optimal price for configuration $e$, and $u_t^{\varsigma}(e,p^*)$ is defined in Equation (\ref{eq:sys:sellerreward}) in Section \ref{sec:sys:trade:seller}. Define $\Delta_{(e,p)}^\varsigma=u_t^{\varsigma}(e^*,p^*) - u_t^{\varsigma}(e,p)$ as the reward gap between price $p$ with a suboptimal $e$ and the optimal price $p^*$ with $e^*$. We establish the following regret bound by leveraging drift-based control techniques for non-stationary rewards \cite{kocsis2006bandit}:

\begin{theorem}[Regret Bound of \textit{CCB} under General Structure]
The cumulative configuration regret incurred by Algorithm~2 over $T_{\varsigma}$ rounds within cluster $\varsigma$ is bounded as follows:

\begin{equation}
 R^{c}_{\varsigma}(T_{\varsigma}) \le
\bar{r}\sum_{e \ne e^*}
\max_{p \in \mathcal{P}}
\left(
\frac{16 \nu^2_\varsigma \log T_{\varsigma}}{(\Delta_{(e,p)}^\varsigma / 2)^2} + 2\omega_\varsigma + \frac{\pi^2_\varsigma}{3}
\right),
\end{equation}

\noindent where $\nu_\varsigma$ is the sub-Gaussian parameter controlling reward variance, $\omega_\varsigma$ bounds the number of biased selections of suboptimal configurations, and $\pi^2_\varsigma / 3$ captures residual exploration.
\end{theorem}

The proof builds on martingale concentration and drift control techniques \cite{kocsis2006bandit}. Despite overlapping reward distributions, the algorithm maintains reliable performance by gradually refining estimates and reducing exploration in low-reward regions. The dominant regret term \( \frac{16 \nu^2_\varsigma \log T_\varsigma}{(\Delta_p^\varsigma / 2)^2} \) reflects the difficulty in distinguishing between near-optimal configurations.

This section presented Algorithm~2, which addresses \emph{Key Question 1} for adaptively selecting retrieval configurations. The chosen configuration $e_t$ is then passed to Stage II for pricing strategy learning, as described in the next section.

\section{Stage II: Pricing Strategy Learning} \label{sec:stage2}

In this section, we present the contextual pricing bandit algorithm (\textit{CPB}) for adaptively selecting prices under variable and complex feedback in Stage II, aiming to minimize the pricing regret $R_\varsigma^p$ in Equation (\ref{eq:framework:decompose}). Section~\ref{sec:stage2:setup} introduces the problem reformulation, Section~\ref{sec:stage2:design} presents the algorithm design, and Section~\ref{sec:stage2:analysis} provides the analysis, showing that \textit{CPB} achieves sublinear regret with polynomial time complexity.

\begin{algorithm}[tb]
\caption{Contextual Pricing Bandit (\textit{CPB})}
\label{alg:pricingucb}
\KwIn{$\varsigma_t,~e_t,~\mathcal{P},~\mathcal{H}$}
\KwData{ taylor order $k$, bias $\Upsilon$, failure probability~$\delta$, smoothness constant $\beta$, reference coefficients ${\eta}$}
Partition $\mathcal{P}=[\underline{p},~\bar{p}]$ into $N$ equal intervals, denoted as $I_j=[a_j,a_{j+1}]$, where $a_j=\underline{p}+j(\overline{p}-\underline{p})/N$\;

\ForEach{interval $j\in\{0, 1, \ldots, N-1\}$}{
    Retrieve rewards $\mathcal{R}_j=\{r_{t'}|\varsigma_{t'}=\varsigma_t, e_{t'} = e_t,j_{t'}=j\}$ from $\mathcal{H}$\;

    \uIf{$\left |\mathcal{R}_j\right |=0$}{
        $\Gamma_j =\infty$
    }
    \Else{
        Compute $\Phi_j = 4\overline{p} \sqrt{2}\kappa\ln(\kappa T+1)\left(\varphi + \frac{\beta+\sqrt{2}}{\left |\mathcal{R}_j\right |}\right)$\;
         Compute $\Gamma_j= \frac{1}{\left|\mathcal{R}_j\right|} \sum_{r\in\mathcal{R}_j} r_t +\Phi_j$\;
    }
}
Compute $j_t = \arg\max_j \Gamma_j$\;

Set $p_t \leftarrow \textit{LAB}(\varsigma_t,e_t,j_t,\mathcal{H})$\;

\KwOut{$(j_t,p_t)$}

\end{algorithm}

\subsection{Problem Reformulation}\label{sec:stage2:setup}
To facilitate effective learning under variable and complex feedback, we begin by reformulating the expected reward expression for the pricing objective in Problem~$\mathbb{P}_1$. While the original form, given in Equation~(\ref{eq:sys:sellerreward}) in Section~\ref{sec:sys:trade:seller}, captures the interaction between cost and utility, its entangled structure complicates direct learning in online settings.

To disentangle these factors and enable more tractable learning, we introduce a reformulated expected reward:

\begin{equation}
u_t^{\varsigma_t}(e_t, p_t) = \chi^{\varsigma_t}(e_t, p_t) \cdot p_t,
\end{equation}

\noindent where $\chi^{\varsigma_t}(e_t, p_t)$ denotes a cost-adjusted response function that accounts for result quality and pricing, serving as a proxy for immediate utility and potential future engagement. This function must be learned online from feedback.

We define a normalized utility variable $y_t = r_t / p_t$, which captures the buyer’s cost-adjusted utility for the transaction. Its expectation under a given configuration and price is modeled by $\chi^{\varsigma_t}(e_t, p_t)$ as follows:

\begin{equation}
\mathbb{E}[y_t \mid e_t, p_t, \varsigma_t] = \chi^{\varsigma_t}(e_t, p_t).
\end{equation}

Accordingly, we redefine the cumulative pricing regret as follows:
\begin{equation}
\label{eq:stage2:regret}
\begin{aligned}
R^p_{\varsigma}(T_{\varsigma}) = \sup_{p^* \in \mathcal{P}}~\sum_{t=1}^{T_{\varsigma}} \chi^{\varsigma_t}(e^*_t, p^*) \cdot p^* - \sum_{t=1}^{T_{\varsigma}} \chi^{\varsigma_t}(e^*_t, p_t) \cdot p_t.
\end{aligned}
\end{equation}

\noindent This expression quantifies the regret of the learned pricing strategy, measured against the reward achieved by the optimal price under the best configuration. To enable theoretically grounded learning of $\chi^{\varsigma}(e, p)$, we introduce the following general smoothness assumption.

\begin{assumption}[Hölder Smoothness]\label{assumption:smooth}
Let $n \in \mathbb{N}$ and $\beta > 0$. The function $\chi^{\varsigma}(e, p)$ lies in the Hölder class $\Sigma^n(\mathcal{E} \times \mathcal{P}; \beta)$ if 
\[
\sup_{(e, p)} \left| \partial^{\bm{\alpha}} \chi^{\varsigma}(e, p) \right| \le \beta, \quad \forall\, \bm{\alpha} \in \mathbb{N}^2,\; |\bm{\alpha}| < n,
\]

\noindent and all $(k-1)$-order partial derivatives are Lipschitz continuous, i.e., $\forall~e,~e'\in\mathcal{E},~ p,~p'\in\mathcal{P},~|\bm{\alpha}| = n-1$: 
\[
\left| \partial^{\bm{\alpha}} \chi^{\varsigma}(e, p) - \partial^{\bm{\alpha}} \chi^{\varsigma}(e', p') \right|
\le \beta \cdot |(e, p) - (e', p')|.
\]
\end{assumption}

The Hölder smoothness assumption characterizes the local regularity of function $\chi^{\varsigma}(e, p)$, which facilitates local approximation and learning under variable and complex feedback.

\subsection{Algorithm Design: CPB}\label{sec:stage2:design}

Algorithm 3 (\textit{CPB}) consists of two key components. The first (Section \ref{sec:stage2:design:interval}) partitions the price range and selects promising intervals by the UCB principle. The second (Section~\ref{sec:stage2:design:local}) refines price selection within the chosen interval via a Local Approximation Bandit (\textit{LAB}) using multivariate Taylor approximation.\footnote{
{A potential extension is to employ Bayesian methods, such as Gaussian Processes }\cite{williams2006gaussian}{ or Bayesian Neural Networks }\cite{neal2012bayesian}{, to model complex feedback and uncertainty. These methods are computationally demanding and difficult to align with regret guarantees, and are therefore left for future work.}
}

\subsubsection{\textbf{Interval Selection}} \label{sec:stage2:design:interval}

We begin with coarse-grained exploration by partitioning $\mathcal{P} = [\underline{p}, \overline{p}]$ into $N$ equal-width sub-intervals (Line~1). Each interval is $I_j = [a_j, a_{j+1}]$, where endpoints are $a_j = \underline{p} + j(\overline{p} - \underline{p}) / N$ for $j \in \{0, 1, \ldots, N-1\}$. 

We then select the most promising interval $j_t$ (Lines~2–9). For each interval $I_j$, we maintain the average reward $\hat{r}_j$ and a confidence term $\Phi_j$ reflecting uncertainty. These are combined into an optimistic estimate $\Gamma_j = \hat{r}_j + \Phi_j$. The interval with the highest $\Gamma_j$ is selected, balancing exploitation and exploration.

The selected interval $I_{j_t}$ is then passed to the local approximation stage, where a Local Approximation Bandit (\textit{LAB}) is applied to refine the pricing decision (Line 10).

\subsubsection{\textbf{Local Taylor Approximation}}\label{sec:stage2:design:local}

We next introduce Algorithm~4 (\textit{LAB}), which determines the final price within the selected interval by Taylor approximation.  By expanding $\chi^\toc(e, p)$ at the anchor $(e_t,a_{j_t})$ using a multivariate Taylor series with respect to the configuration offset $\eta = e -e_t$ and price offset $\Delta = p - a_{j_t}$, we transform the original pricing problem into a locally linear regression task. This formulation enables efficient estimation of response using a locally expressive and statistically stable feature representation.

\begin{algorithm}[tb]
\caption{Local Approximation Bandit (\textit{LAB})}
\KwIn{ $\varsigma_t$, $e_t$, $j_t$, $\mathcal{H}$, }
\KwData{ Taylor order $n$, error constant $\Upsilon$, smoothness constant $\beta$, failure probability~$\delta$, offset ${\eta}$}

Retrieve pairs $\mathcal{D}=\{(p_{t'},y_{t'})|\varsigma_{t'}=\varsigma_t, e_{t'} = e_t,j_{t'}=j\}$ from $\mathcal{H}$\;

Compute feature vector
\[        \phi(e,p) = \left\{ \eta^{i_{e}}\Delta^{i_p} \right\}_{i_{e} + i_p < n}, \quad i_e, i_p \in \mathbb{N};
\]

Compute matrix $\Lambda = I_{\kappa \times \kappa} + \sum_{(p,y) \in \mathcal{D}} \phi(e,p) \phi(e,p)^\top$\;

Compute ridge regression:
\[
\hat{\theta}^\toc = \arg\min_{\theta \in \mathbb{R}^\kappa} \sum_{(p,y) \in \mathcal{D}} \left(y - \langle \theta, \phi(\epf,p) \rangle \right)^2 + \|\theta\|_2^2;
\]

Set $\rho = \beta\sqrt{\kappa} + \Upsilon\sqrt{|\mathcal{D}|} + \sqrt{2\kappa\ln(4\kappa |\mathcal{D}|/\delta)} + 2$\;
Compute $p_t= \arg\max_{p \in I_{j_t}} \; p \cdot \min(1,\langle \hat{\theta}^\toc, \phi(e,p) \rangle  + \rho  \sqrt{\phi(e,p)^\top \Lambda^{-1} \phi(e,p)} + \Upsilon)$\;
\KwOut{$p_t$}
\end{algorithm}

We define a feature vector $\phi(e,p)$ that includes all monomials of total degree up to $n - 1$ in $(\eta, \Delta)$:

\begin{equation}
    \phi(e,p) = \left\{ \eta^{i_{e}}\Delta^{i_p} \right\}_{i_{e} + i_p < n}, \quad i_e, i_p \in \mathbb{N}.
\end{equation}

\noindent This expansion captures all polynomial interactions between configuration deviation and price shift up to order $n-1$. The resulting feature vector $\phi_e(p)$ has dimension $    \kappa = \frac{n(n+1)}{2}$, since it includes all combinations of non-negative integer powers $(i_{\eta}, i_p)$ satisfying $i_{\eta} + i_p < n$. 

Accordingly, we approximate the response function as:

\begin{equation}
    \chi^\toc (e, p) \approx \langle \theta^\toc, \phi(e,p) \rangle + m_t,
\end{equation}

\noindent where $\theta^\toc \in \mathbb{R}^{\kappa}$ is the local coefficient vector to be learned, and $m_t$ is a bias variable.

The algorithm collects historical pairs $\mathcal{D}$ (Line 1) with the same $(\varsigma_t, e_t, j_t)$ and fits $\theta^\varsigma$ by ridge regression (Lines 2-4). Finally, the algorithm selects price $p_t$ by maximizing the optimistic estimate of the expected reward (Lines 5-6).

\subsection{Theoretical Analysis}\label{sec:stage2:analysis}

We first analyze the theoretical bound of Algorithms 3 and 4. We then present the overall time complexity of the complete framework, building on the descriptions of all algorithms.

\subsubsection{\textbf{Taylor Approximation Guarantee}}
To assess the accuracy of local polynomial modeling, we begin by analyzing the approximation error and present the following lemma:

\begin{lemma}[Approximation Error Bound]
Assume $\chi^\varsigma(\epf, p)\in \Sigma^n(\mathcal{E} \times \mathcal{P}; \beta)$ (as defined in Assumption 1). Let $B = \Xi \times I$ be a local region centered at $(e_t, a_{j_t})$, where $|\Xi| = \eta$ and $|I| = (\underline{p}+\bar{p})/N$. Then, the local approximation error is bounded by:

\begin{equation}
    \Upsilon = \frac{\beta n}{(n-1)!}\left(\eta + (\underline{p}+\bar{p})/N\right)^n.
    \label{eq:stage2:bias}
\end{equation}
\end{lemma}

The proof of Lemma 1 is given in Appendix B in the technical report \cite{report}. This lemma shows that the Taylor approximation error decays polynomially as the local region shrinks, validating the use of low-degree polynomials.

\subsubsection{\textbf{Regret Bound of Local Approximation}}
We next analyze the regret of Algorithm~4 (\textit{LAB}) within a particular interval $I$.
\begin{lemma} [Regret Bound of \textit{LAB}]
Fix a particular $I$, and let $\hat{p}_1, \ldots, \hat{p}_t$ be the prices selected in $t$ rounds of Algorithm 4 within $I$. With probability at least $1 - \mathcal{O}(1/{T_\varsigma^3})$, the average pricing regret per round in Algorithm 4 satisfies:
\begin{equation}
    \begin{aligned}
        & \frac{1}{t} \sum_{\tau \leq t} \left[ \max_{p \in I} p \cdot\chi^\varsigma(e,p) - \hat{p}_\tau\cdot \chi^\varsigma(e,\hat{p}_\tau) \right] \\
 & \leq    2\overline{p}\sqrt{2}\kappa\ln(\kappa T_\varsigma+1)\left(\Upsilon +\frac{\beta+\sqrt{2}}{\sqrt{t}}\right).
    \end{aligned}
\end{equation}
\label{lm:priceregret2}
\end{lemma}

The proof of Lemma 2 is given in Appendix C in the technical report \cite{report}. This bound indicates that the accuracy of local learning improves as the number of samples increases since the second term vanishes with growing $t$.

\subsubsection{\textbf{Regret Bound of Pricing Learning}}
Finally, we establish a bound on the cumulative pricing regret of Algorithm 3 (\textit{CPB}) within a cluster, as stated in the following theorem.
\begin{theorem}[Regret Bound of \textit{CPB}]
\label{th:pricereg-epf}
Assume $\chi^\varsigma(e, p) \in \Sigma^{n}(\mathcal{E} \times \mathcal{P}; \beta)$ with $\beta > 0$, and let $N = \lceil T_\varsigma^{1/(2n+1)} \rceil$. Then, with probability at least $1 - \mathcal{O}(1/T_\varsigma)$, the cumulative pricing regret of Algorithm~3 is bounded by:
\begin{equation}
\begin{aligned}
R^{p}_\varsigma(T_\varsigma) \le & 6\sqrt{2}\kappa T_\varsigma^{\frac{n+1}{2n+1}}\ln(\kappa T_\varsigma+1) \left( \beta + 2(\Upsilon + \sqrt{2})  \right).
\end{aligned}
\end{equation}
\end{theorem}

The proof of Theorem 3 is provided in Appendix D of the technical report \cite{report}. This bound confirms that the regret grows sublinearly with $T_\varsigma$, ensuring that the average per-round regret diminishes over time. As $n \to \infty$, the exponent $\frac{n+1}{2n+1}$ approaches $1/2$, yielding an asymptotic regret of $\tilde{O}(\sqrt{T_\varsigma})$.\footnote{
{Our regret guarantees do not rely on specific distributions or modalities and hold as long as queries are represented as vectors, making the framework applicable to heterogeneous traffic and diverse data types.}
}

\subsubsection{\textbf{Time Complexity}} With all algorithmic components introduced, we proceed to analyze the overall time complexity.

\begin{theorem}[Overall Time Complexity of \textit{VTHB}]
The total time complexity of the Algorithm 1 over $T$ rounds is
\[
\mathcal{O}\left(T\left(|\mathcal{E}| + N + \kappa^2\right)\right),
\]
where $|\mathcal{E}|$ is the number of candidate configurations, $N$ is the number of price intervals, and $\kappa$ is the feature dimension
\label{thm:time}
\end{theorem}

The proof of Theorem~\ref{thm:time} is provided in Appendix E of the technical report \cite{report}. This result establishes that the framework admits polynomial time complexity and is practically efficient, using incremental updates of historical statistics in \textit{CCB} and \textit{CPB}, as well as fast matrix inverse updates in \textit{LAB} (e.g., using rank-one Sherman–Morrison or Cholesky decomposition).

This section proposed and analyzed \textit{CPB}, a contextual pricing learner that integrates interval-based selection and Taylor approximation, addressing \emph{Key Question~2} on learning pricing strategies under variable and complex feedback. We next proceed to the experimental evaluation to validate the effectiveness of the overall framework.

\section{Experimental Evaluation}\label{sec:experiments}
In this section, we conduct comprehensive experiments on four real-world datasets to evaluate our framework's performance. Section \ref{sec:exp:setting} describes the experimental settings, and Section \ref{sec:exp:result} presents a series of experimental result analyses and corresponding insights.

\subsection{Experimental Settings} \label{sec:exp:setting}
\subsubsection{\textbf{Datasets}}

Table~\ref{tab:datasets} summarizes the four real-world datasets used in our experiments, covering diverse modalities.

\begin{table}[ht]
\centering
\caption{Description of experimental datasets.}
\label{tab:datasets}
\begin{tabular}{p{1.6cm} p{0.9cm} p{5cm}}
\toprule
\textbf{Dataset} & \textbf{Modality} & \textbf{Description} \\
\midrule
\gist~\cite{jegou2010product} & Visual & 1M 960-dimensional vectors for content-based image retrieval. \\
\sift~\cite{jegou2010product} & Visual & 1M 128-dimensional SIFT descriptors used in image matching. \\
\msong~\cite{Bertin-Mahieux2011} & Auditory & 1M audio feature vectors from the Million Song Dataset based on timbre and chroma descriptors. \\
\glove~\cite{pennington2014glove} & Textual & 1.2M 100-dimensional word embeddings trained using the GloVe algorithm. \\
\bottomrule
\end{tabular}

\end{table}

{For each dataset, we use 90\% of the vectors to build the ANN index and reserve 10\% as queries. 
We construct each buyer query as $\bm{q}_t=(\bm{v}_t,c_t,k_t)$, where $\bm{v}_t$ comes from the dataset and $c_t,k_t$ are sampled from uniform distributions over predefined ranges. 
Buyer responses combine a price-sensitive bimodal base-demand function with a quality factor that increases with retrieval quality. This simulation provides a controllable environment with known ground truth, enabling fair comparisons when counterfactual labels are unavailable.
}

\subsubsection{\textbf{Baselines}}
Given the absence of existing methods for vector data trading, we construct representative baselines by separately evaluating retrieval configuration and pricing. For retrieval, we include: (a) a static configuration method (STCF), which fixes retrieval parameters and reflects conventional non-adaptive systems~\cite{malkov2018efficient}; and (b) a random configuration method (RDCF), which selects configurations uniformly at random as a naive reference. For pricing, we consider: (c) static pricing (STP), which posts a fixed price; (d) random pricing (RDP), which samples uniformly from a price range; (e) linear pricing (LinP), which fits a linear response curve~\cite{li2010contextual}; and (f) convex pricing (ConP), which models smoother but non-linear responses~\cite{suggala2024second}. Together, these baselines span simple heuristics to parametric estimators and serve as comparative references for evaluating the effectiveness of our hierarchical framework (VTHB) under coupled retrieval–pricing dynamics.

\subsubsection{\textbf{Evaluation Metrics}}
We adopt two key metrics to evaluate performance. The cumulative regret measures the total loss compared to an ideal strategy that always chooses the best configuration and price, indicating how effectively the method learns over time. The average reward reflects the utility gained per round and captures the trade-off between retrieval cost and buyer satisfaction under varying conditions.

\subsection{Result Analysis}  \label{sec:exp:result}
In this subsection, we present the experimental results in two parts: overall performance (Section~\ref{sec:experiments:overall}) and parameter study (Section~\ref{sec:experiments:parameter}).

\subsubsection{\textbf{Overall Performance}}\label{sec:experiments:overall}

\begin{figure}[t]
    \centering
    \includegraphics[width=0.45\textwidth]{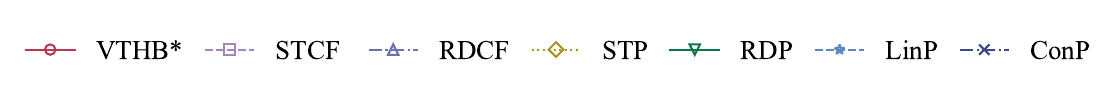} \\
    \subfloat[\gist]{\label{fig:exp:gist:regret}\includegraphics[width=0.23\textwidth]{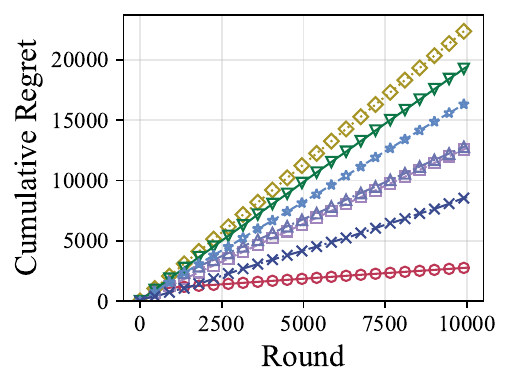}}
    \subfloat[\sift]{\label{fig:exp:sift:regret}\includegraphics[width=0.23\textwidth]{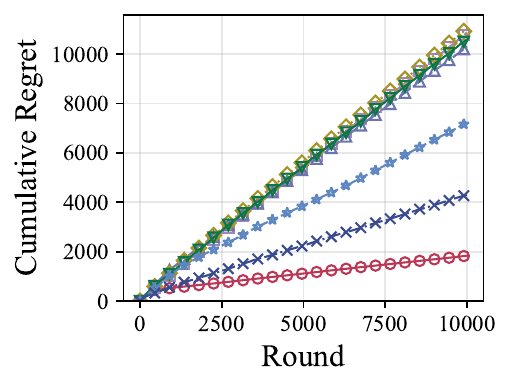}} \\
    \subfloat[\msong]{\label{fig:exp:msong:regret}\includegraphics[width=0.23\textwidth]{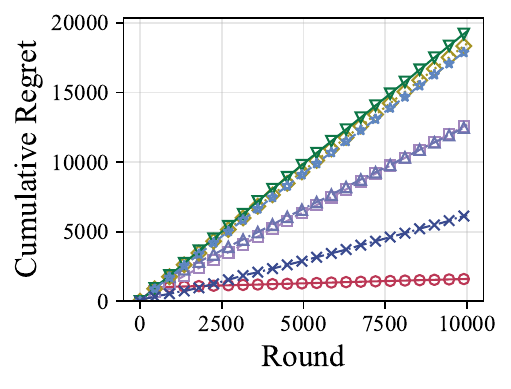}}
    \subfloat[\glove]{\label{fig:exp:glove:regret}\includegraphics[width=0.23\textwidth]{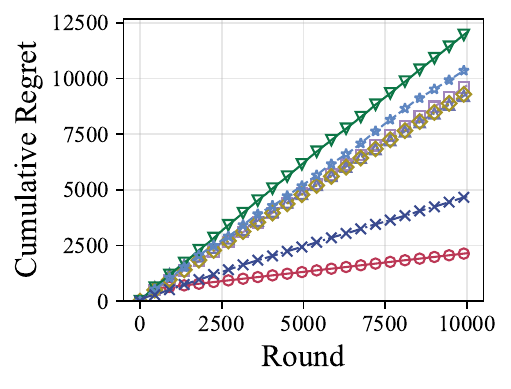}}
    \caption{Comparison of methods on cumulative regret.}
    \label{fig:exp:regret}
    
\end{figure}

\begin{figure}[hbpt]
    \centering
    \includegraphics[width=0.45\textwidth]{Figure/experiments/glove_results_legend.pdf} \\
    \subfloat[\gist]{\label{fig:exp:gist:reward}\includegraphics[width=0.23\textwidth]{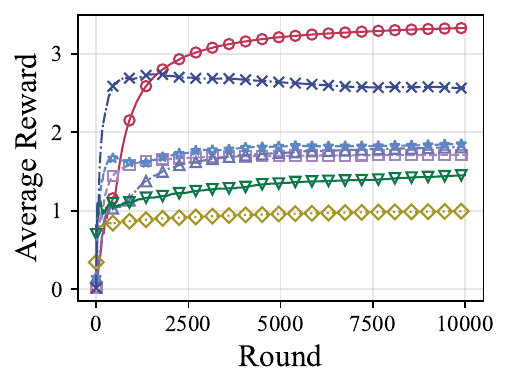}}
    \subfloat[\sift]{\label{fig:exp:sift:revenue}\includegraphics[width=0.23\textwidth]{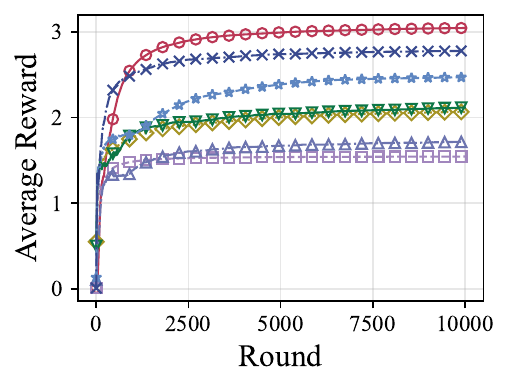}} \\
    \subfloat[\msong]{\label{fig:exp:msong:revenue}\includegraphics[width=0.23\textwidth]{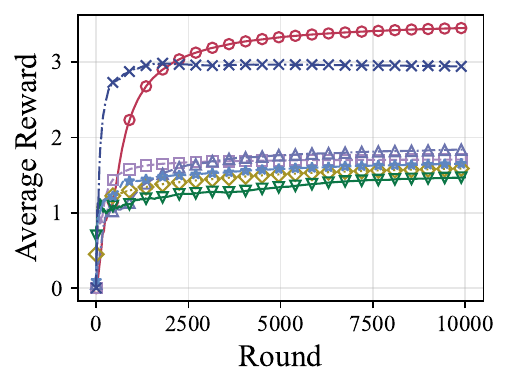}}
    \subfloat[\glove]{\label{fig:exp:glove:revenue}\includegraphics[width=0.23\textwidth]{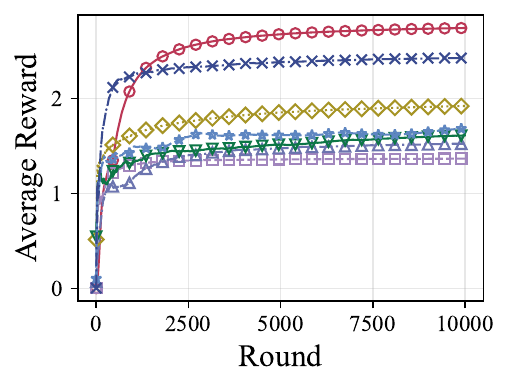}}
    \caption{Comparison of methods on average reward.}
    \label{fig:exp:reward}
    
\end{figure}

We evaluate all methods over 10{,}000 rounds on four datasets. Figs.~\ref{fig:exp:regret} and \ref{fig:exp:reward} present cumulative regret and average reward, reflecting learning efficiency and economic performance. Across all datasets, our method outperforms baselines, achieving fast regret reduction and sustained rewards, consistent with the sublinear regret bound.

For cumulative regret, our method attains the lowest values: 2767.64 on GIST1M, 1836.59 on SIFT1M, 1604.31 on MSONG, and 2155.48 on GLOVE1.2M. Compared with the best baseline (ConP), this represents reductions of 67.9\%, 57.4\%, 74.0\%, and 47.3\%, respectively, and over 85\% versus static pricing (STP) on GIST1M, MSONG, and GLOVE1.2M. These results demonstrate the effectiveness of adaptive configuration learning with contextual clustering and UCB-guided exploration. For average reward, our method also leads, reaching 3.3288 on GIST1M, 3.0470 on SIFT1M, 3.4497 on MSONG, and 2.7418 on GLOVE1.2M, improvements of 29.9\%, 9.6\%, 17.3\%, and 12.9\% over ConP. While ConP often converges prematurely due to convex modeling, our method continues to refine pricing via cluster-specific learning and localized exploration, yielding sustained reward gains. This highlights the necessity of contextual, adaptive learning for long-term optimization in dynamic trading environments.

\subsubsection{\textbf{Parameter Study}}\label{sec:experiments:parameter}

To evaluate the robustness of our method, we conduct a parameter study on both retrieval configuration and pricing performance, showing that it maintains stable and adaptive behavior across different datasets.

\emph{(a) Configuration Performance:} To assess configuration effectiveness under price constraints, we compare VTHB with RDCF and STCF while varying the price upper bound $\bar{p}$ (Table~\ref{tab:avg_reward_bar_p}). These baselines focus only on retrieval configuration, isolating Stage I performance in our framework. Across all datasets and $\bar{p}$ values, VTHB consistently outperforms both, with RDCF second and STCF weakest under tight budgets. As $\bar{p}$ rises from 5 to 10, all methods improve notably; for example, VTHB’s reward increases from 0.8042 to 3.4098 on \textsc{GIST} and from 1.0491 to 3.4014 on \textsc{MSONG}, reflecting greater flexibility in configuration choices.

Performance generally peaks at $\bar{p}=10$, after which rewards decline, consistent with the nonlinear and sparse structure of the response function at higher prices. Excessive exploration under large $\bar{p}$ yields less informative feedback, leading to suboptimal configuration–price pairs. Thus, moderately bounded price spaces are essential for stable and efficient learning. Notably, at $\bar{p}=10$, VTHB exceeds RDCF and STCF by over 86\% in most cases and retains a clear lead even at $\bar{p}=20$, confirming its robustness to price variation and adaptability across scenarios.

\begin{table}[tb]
\centering
\caption{Average reward under varying values of $\bar{p}$.}
\label{tab:avg_reward_bar_p}
\begin{tabular}{c l c c c c}
\toprule
\textbf{$\bar{p}$} & \textbf{Method} & \textbf{\gist} & \textbf{\sift} & \textbf{\glove} & \textbf{\msong} \\
\midrule
\multirow{3}{*}{5} 
  & VTHB* & 0.8042 & 2.2144 & 1.7381 & 1.0491  \\
  & RDCF  & 0.6596 & 1.7481 &  1.1457 & 0.7548  \\
  & STCF  & 0.5780 & 1.5979 & 1.0428 & 0.6812 \\
\midrule
\multirow{3}{*}{10} 
  & VTHB* & 3.4098  & 3.1579 & 2.8380 & 3.4014 \\
  & RDCF  & 1.8335  & 1.7639 & 1.5509 & 1.8253 \\
  & STCF  & 1.7510 & 1.5910 & 1.3581 & 1.7463  \\
\midrule
\multirow{3}{*}{15} 
  & VTHB* & 1.8865 & 2.4584 & 1.9823 & 1.7160 \\
  & RDCF  & 1.4851 & 1.5917 & 1.3350 & 1.4672 \\
  & STCF  & 1.3701 & 1.4788 & 1.2149 & 1.3537 \\
\midrule
\multirow{3}{*}{20} 
  & VTHB* & 1.5402 & 2.1803 & 1.3714 & 1.4079 \\
  & RDCF  & 1.3856 & 1.6828 & 1.3161 & 1.3644 \\
  & STCF  & 1.2797 & 1.5622 & 1.2162 & 1.2600 \\
\bottomrule
\end{tabular}

\end{table}

\begin{figure}[hbpt]
    \centering
    \subfloat[\gist]{\label{fig:exp:gist:k}\includegraphics[width=0.23\textwidth]{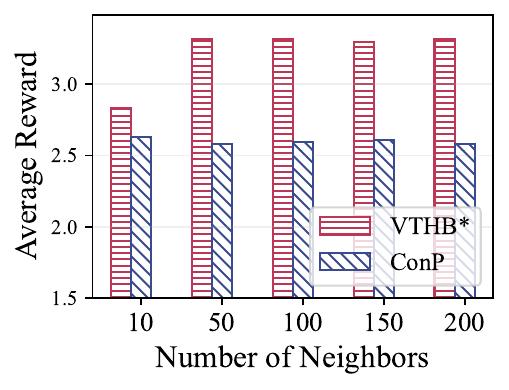}}
    \subfloat[\sift]{\label{fig:exp:sift:k}\includegraphics[width=0.23\textwidth]{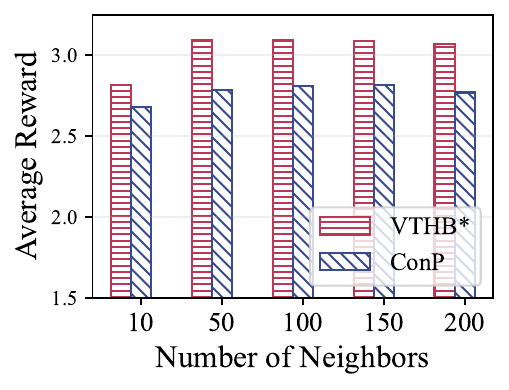}} \\
    
    \subfloat[\msong]{\label{fig:exp:msong:k}\includegraphics[width=0.23\textwidth]{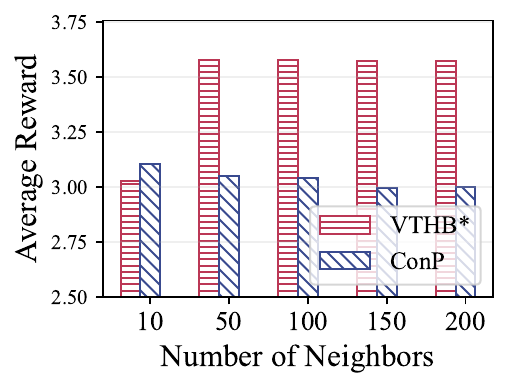}}
    \subfloat[\glove]{\label{fig:exp:glove:k}\includegraphics[width=0.23\textwidth]{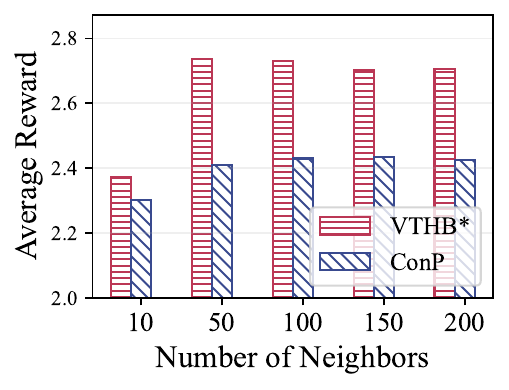}}
    \caption{Impact of the number of neighbors.}
    \label{fig:exp:k}
    
\end{figure}

\emph{(b) Pricing Performance} 
We evaluate pricing performance by examining the impact of the number of neighbors ($k$) on average reward across four datasets, as shown in Fig. \ref{fig:exp:k}, comparing our method (VTHB) with the best-performing baseline (ConP). Overall, VTHB consistently outperforms ConP across all settings. For instance, on GIST1M (Fig. \ref{fig:exp:gist:k}), VTHB achieves up to 28.5\% higher reward at $k=50$ and remains stable as $k$ increases. On SIFT1M (Fig. \ref{fig:exp:sift:k}), it maintains a consistent lead with an 11.2\% gain at $k=50$. The performance gap becomes more pronounced on MSONG and GLOVE1.2M (Figs. \ref{fig:exp:msong:k} and~\ref{fig:exp:glove:k}), where VTHB exceeds ConP by 19.1\% and 11.6\% at $k=200$, respectively. These results highlight VTHB’s ability to leverage richer neighborhood information to support effective pricing without overfitting.

Notably, ConP’s performance fluctuates or degrades as $k$ increases, particularly on MSONG and GLOVE1.2M, likely due to the inclusion of irrelevant neighbors. In contrast, VTHB maintains a stable upward trend, underscoring its robustness to hyperparameter choices. This demonstrates that VTHB not only achieves higher pricing performance but also offers greater resilience to parameter variations, which stems from its hierarchical structure and adaptive learning mechanism.

\begin{figure}[hbpt]
    \centering
    \subfloat[GIST1M]{\includegraphics[width=0.23\textwidth]{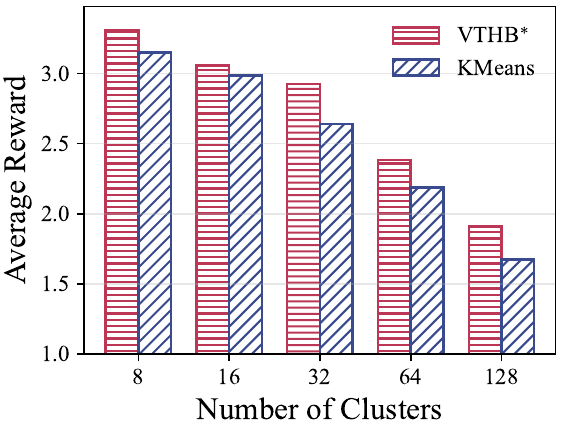}}
    \subfloat[SIFT1M]{\includegraphics[width=0.23\textwidth]{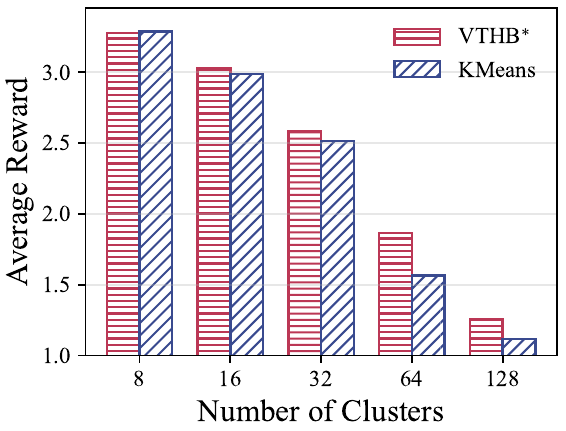}} \\
    \subfloat[MSONG]{\includegraphics[width=0.23\textwidth]{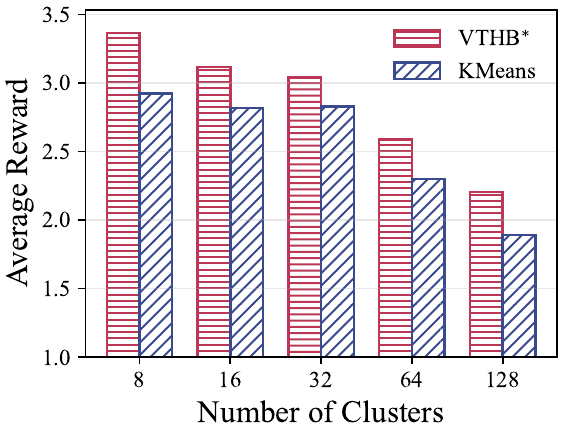}}
    \subfloat[GLOVE1.2M]{\includegraphics[width=0.23\textwidth]{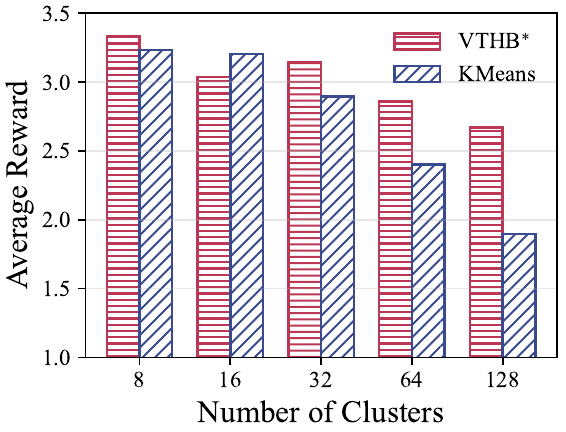}}
    \caption{{Reward sensitivity across cluster granularities.}}
    
    \label{fig:exp:add:reward1}
\end{figure}

\begin{figure}[hbpt]
    \centering
    \subfloat[GIST1M]{\includegraphics[width=0.23\textwidth]{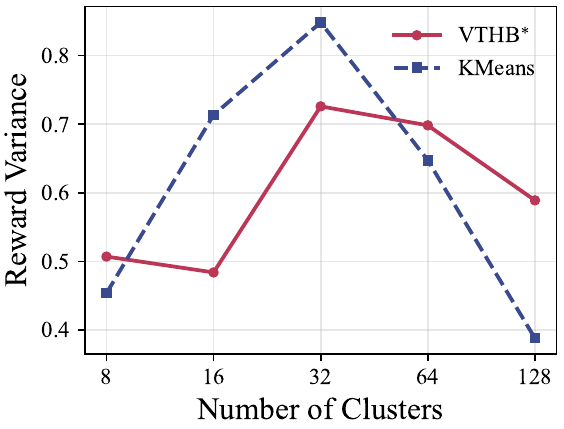}}
    \subfloat[SIFT1M]{\includegraphics[width=0.23\textwidth]{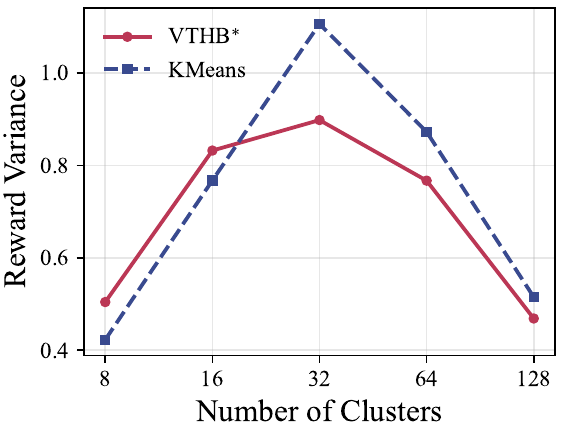}} \\
    \subfloat[MSONG]{\includegraphics[width=0.23\textwidth]{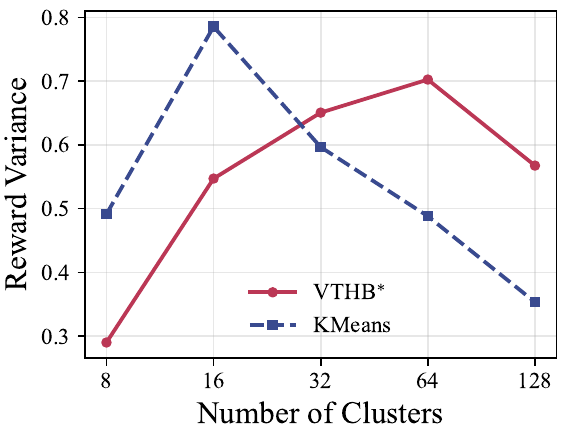}} 
    \subfloat[GLOVE1.2M]{\includegraphics[width=0.23\textwidth]{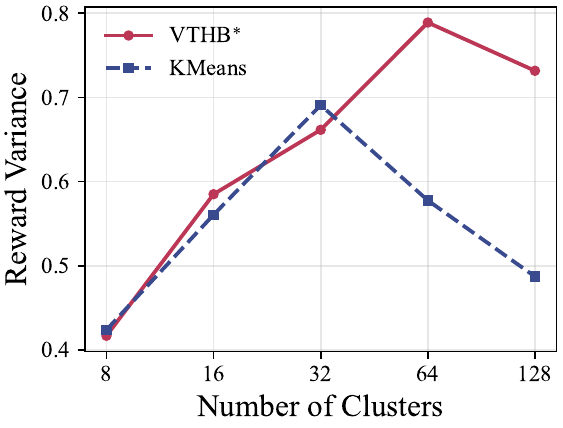}}
    \caption{{Intra-cluster reward variance across cluster granularities.}}
    
    \label{fig:exp:add:var1}
\end{figure}

\emph{(c) Clustering Performance}
{As shown in Fig.~}\ref{fig:exp:add:reward1}{, we evaluate reward sensitivity with respect to the number of clusters on GIST1M, SIFT1M, MSONG, and GLOVE1.2M, comparing VTHB with KMeans. VTHB consistently achieves comparable or higher rewards, with performance most stable in the range of 16–32 clusters. For example, on GIST1M at 32 clusters, VTHB delivers about 20\% higher reward than KMeans. When the number of clusters is too small, heterogeneous queries are merged and adaptation limited; when too large, fragmentation reduces per-cluster sample size. Overall, VTHB exhibits more graceful degradation than KMeans as granularity increases, demonstrating stronger robustness under fine partitions.}

{As shown in Fig.~}\ref{fig:exp:add:var1}{, we analyze intra-cluster reward variance as a measure of homogeneity. On GIST1M and SIFT1M, VTHB yields overall lower variance than KMeans---for example, about 25\% lower on SIFT1M at 32 clusters---indicating that the three-factor clustering forms more coherent cost--quality--response relations at coarse-to-moderate granularities. On MSONG and GLOVE1.2M, the variance of VTHB is sometimes higher, but its average reward remains higher (see Fig.~}\ref{fig:exp:add:reward1}{), suggesting that variance alone does not fully determine performance. Instead, by explicitly incorporating $(c_t,k_t)$, VTHB better aligns clustering with downstream learning objectives, leading to more stable reward outcomes even when variance is not strictly minimized.}

In this section, we showed that the proposed framework performs better than baseline methods in both regret and reward. The results confirm the framework's effectiveness and robustness across different datasets and settings.

\section{Conclusion}\label{sec:conclusion}

In this work, we study vector data trading and propose a two-stage hierarchical bandit framework that jointly optimizes retrieval configurations and dynamic pricing. Theoretically, we establish sublinear regret bounds for both stages and show that the algorithm runs in polynomial time. Experiments on four real-world datasets confirm that our method consistently outperforms baseline approaches in cumulative reward and regret, highlighting its value in real-world vector data markets.

Future work will focus on several promising directions.
{One extension is to enhance robustness against adversarial or strategic feedback, for example, by incorporating adversarial bandit methods or mechanism design tools, so that guarantees hold even when buyer responses are manipulated.}
{We also plan to enrich the trading protocols by considering budget-constrained queries, pre-agreed pricing quotes, and more flexible data protocols.}
{It is also valuable to extend the framework to explicitly model buyer-side decision-making in competitive markets.}
{Finally, to improve sample efficiency in low-traffic clusters, a natural extension is to apply meta-learning or multi-task bandits for cross-cluster knowledge transfer.}

\bibliographystyle{ieeetr}
\bibliography{IEEEabrv,reference}
\end{document}